\documentclass[11pt]{article}

\usepackage[left=1in,right=1in,bottom=1in,top=1in]{geometry}
\usepackage{amsmath,amssymb}
\usepackage{hyperref}
\hypersetup{colorlinks=true, linkcolor=black, citecolor=black, urlcolor=cyan}
\usepackage{graphicx}
\graphicspath{{}}
\usepackage[space]{grffile}
\usepackage{tikz}
\usetikzlibrary{shapes.geometric, arrows}
\tikzstyle{arrow} = [thick,->,>=stealth]
\usepackage{natbib}
\setcitestyle{notesep={; }}

\renewcommand{\abstract}[1]{
 \centerline{
 \begin{minipage}{0.7\linewidth}
 \hrule
 \vskip 0.1in
  \begin{center}
    {\bf Abstract}
  \end{center}
  #1
 \vskip 0.1in
 \hrule
 \end{minipage}}
 \vskip 0.3in}

\usepackage{amsthm}
\newtheorem{theorem}{Theorem}
\newtheorem*{theorem*}{Theorem}

\providecommand{\x}{\mathbf{x}}
\providecommand{\X}{\mathbf{X}}
\providecommand{\I}{\mathbf{I}}
\providecommand{\W}{\mathbf{W}}
\renewcommand{\u}{\mathbf{u}}
\providecommand{\y}{\mathbf{y}}

\providecommand{\yt}{\tilde{\mathbf{y}}}
\providecommand{\rj}{\mathbf{r}_j}
\providecommand{\bb}{\boldsymbol{\beta}}
\providecommand{\bh}{\hat{\beta}}
\providecommand{\bbh}{\hat{\boldsymbol{\beta}}}
\providecommand{\bbt}{\tilde{\boldsymbol{\beta}}}
\providecommand{\be}{\boldsymbol{\eta}}

\providecommand{\bet}{\tilde{\boldsymbol{\eta}}}
\providecommand{\bep}{\boldsymbol{\epsilon}}

\providecommand{\lam}{\lambda}

\providecommand{\cM}{\mathcal{M}}

\providecommand{\cZ}{\mathcal{Z}}
\providecommand{\CV}{\textrm{CV}}
\providecommand{\SE}{\textrm{SE}}

\providecommand{\Fdr}{\textrm{Fdr}}

\providecommand{\fdr}{\textrm{fdr}}

\providecommand{\mFDR}{\textrm{mFdr}}
\providecommand{\mfdr}{\textrm{mfdr}}

\providecommand{\abs}[1]{\left\lvert#1\right\rvert}
\providecommand{\al}[2]{\begin{align}\label{#1}#2\end{align}}

\providecommand{\inD}{\overset{d}{\longrightarrow}}
\newcommand\independent{\protect\mathpalette{\protect\independenT}{\perp}}
\def\independenT#1#2{\mathrel{\rlap{$#1#2$}\mkern2mu{#1#2}}}

\providecommand{\Ex}{\textrm{E}}
\DeclareMathOperator{\EX}{\mathbb{E}}
\providecommand{\Norm}{\textrm{N}}

\title{Feature-specific inference for penalized regression using local false discovery rates}
\author{Ryan Miller\\Department of Mathematics and Statistics\\Grinnell College
  \and
  Patrick Breheny\\Department of Biostatistics\\University of Iowa}
\date{\today}

\begin{document}

\maketitle

\abstract{Penalized regression methods, most notably the lasso, are a popular approach to analyzing high-dimensional data. An attractive property of the lasso is that it naturally performs variable selection.  An important area of concern, however, is the reliability of these selections.  Motivated by local false discovery rate methodology from the large-scale hypothesis testing literature, we propose a method for calculating a local false discovery rate for each variable under consideration by the lasso model.  These rates can be used to assess the reliability of an individual feature, or to estimate the model's overall false discovery rate.  The method can be used for all values of $\lambda$.  This is particularly useful for models with a few highly significant features but a high overall Fdr, a relatively common occurrence when using cross validation to select $\lambda$.  It is also flexible enough to be applied to many varieties of penalized likelihoods including GLM and Cox models, and a variety of penalties, including MCP and SCAD.  We demonstrate the validity of this approach and contrast it with other inferential methods for penalized regression as well as with local false discovery rates for univariate hypothesis tests.  Finally, we show the practical utility of our method by applying it to two case studies involving high dimensional genetic data.}

\section{Introduction}

In recent years, data involving large numbers of features have become increasingly prevalent. Broadly speaking, there are two main approaches to analyzing such data: \textit{large-scale testing} and \textit{regression modeling}. The former entails conducting separate tests for each feature, while the later considers all features simultaneously in a single model.
A major advance in large-scale testing has been the development of methods for estimating \textit{local false discovery rates}, which provide an assessment of the significance of individual features while controlling the false discovery rate across the multiple tests.
We present here an approach for extending local false discovery rates to penalized regression models such as the lasso, thereby quantifying each feature's importance in a way that has been absent in the field of penalized regression until now.

\begin{figure}[htb]
\centering
\includegraphics[width=.95\textwidth]{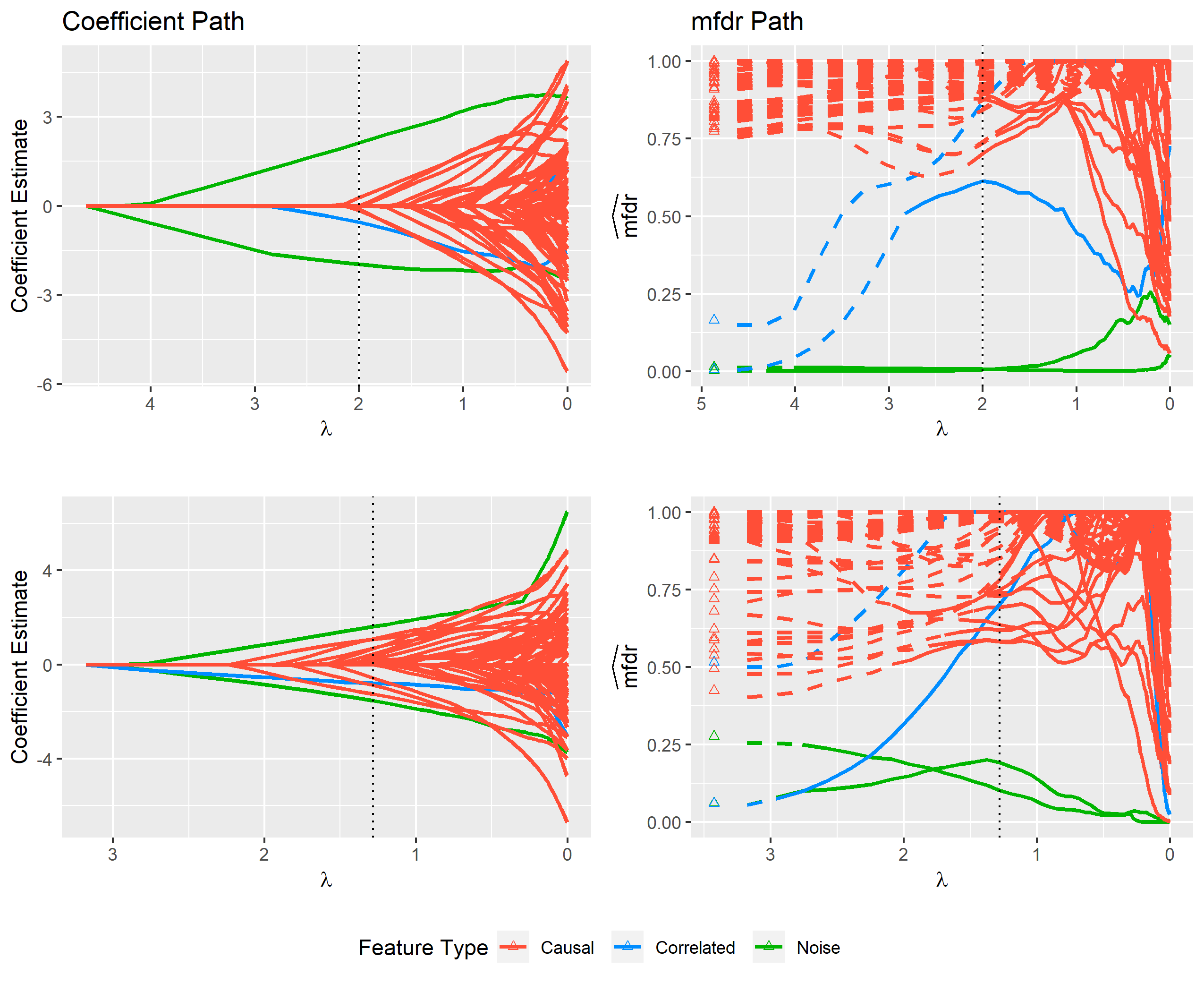}
\caption{\label{Fig:Example} The rows of this figure correspond to two simulated datasets. The column on the left shows the usual lasso coefficient path, while the column on the right displays our method's feature specific local false discovery rates ($\mfdr$) along the lasso path. The triangles prior to the start of the mfdr path are the traditional local false discovery rate estimates resulting from large-scale testing. Along the mfdr path dashed lines indicate the portion of the path where a feature is inactive ($\bh = 0$) in the model. The vertical dotted line shows the value of the penalty parameter, $\lambda$, chosen by cross validation.}
\end{figure}

Figure~\ref{Fig:Example} shows two simulated examples to illustrate the information provided by local false discovery rates in the penalized regression context. In both examples, there are two features causally related to the outcome, two features that are correlated with the causal variables, and 96 features that are purely noise. The panels on the left show the lasso coefficient estimates returned by standard software packages such as \texttt{glmnet} \citep{Friedman2010}, while the panels on the right display the local false discovery rate, with a dotted line at $\lambda_{\CV}$, the point along the path which minimizes cross validation error. In the upper left panel, the model at $\lambda_{\CV}$ contains several noise variables, but the two causal features clearly stand out from others in the coefficient path. The $\mfdr$ plot in the upper right panel confirms this visual assessment -- the two causal features have much lower false discovery rates than the other features. The dataset in the second row demonstrates a more challenging case. Here, it is not obvious from the coefficient path which features are significant and which may be noise. The $\mfdr$ plot in the lower right panel lends clarity to this situation, showing at $\lambda_{\CV}$ that the truly causal features are considerably less likely to be false discoveries.

The two $\mfdr$ paths of Figure~\ref{Fig:Example} also illustrate the connection between the $\mfdr$ approach and traditional large-scale testing approach to local false discovery rates. These univariate local false discovery rates are denoted by the triangles prior to the start of the $\mfdr$ path, and are equivalent to the $\mfdr$ estimates at the beginning of the $\mfdr$ path when no features are active in the model. Initially, each method identifies both causal features, along with some of the correlated features, as important. However as $\lambda$ decreases, and the causal features become active in the model, the regression-based $\mfdr$ method reveals that the correlated features are only indirectly related to the outcome.

Having presented an initial case for the utility of $\mfdr$ and an illustration of the connections it shares with both traditional local false discovery rates and lasso regression, we structure the remainder of the paper as follows: Section~\ref{Sec:mfdr_back} gives a more formal introduction to false discovery rate approaches in the context of both large-scale testing and model based approaches to high dimensional data. Section~\ref{Sec:method} introduces our lasso based $\mfdr$ estimator in the linear regression setting and extends the approach to a more general class of penalized likelihood models including penalized logistic and Cox regression models. Section~\ref{Sec:loc_sims} studies the $\mfdr$ approach using simulation, comparing it to existing methods commonly used in high-dimensional analysis, and Section~\ref{Sec:loc_case} explores two real data case studies where the method proves to be useful.

\section{Background}
\label{Sec:mfdr_back}

In the context of both large-scale testing and model-based approaches, this paper will focus on false discovery rates, a common approach to inference in high-dimensional data analysis. There are two main types of false discovery rates: tail-area approaches, which describe the expected rate of false discoveries for all features beyond a given threshold, and local approaches, which describe the density of false discoveries at a specific point. We adopt the general convention throughout, used by many other authors, of using Fdr to refer to tail-area approaches, and fdr to refer to local approaches, reflecting the traditional use of $F$ and $f$ to refer to distribution and density functions. Both Fdr and fdr have been well studied in the realm of large-scale testing. The seminal Fdr procedure \citep{BH_1995} remains a widely popular approach to Fdr and has led to many extensions and related approaches \citep[][and many others]{Storey2004, Genovese2004, Strimmer2008}. Using an empirical Bayes framework, \citet{Efron2001} introduced the idea of local false discovery rates, a proposal which has also been extended in many ways \citep[e.g.,][]{Muralidharan2010,Stephens2017}. In this section, we provide a brief overview of false discovery rate estimation from an empirical Bayes estimation perspective; for more information, including other ways of motivating FDR estimation and control, see \citet{Farcomeni2008} and \citet{Efron_book}.

Recently, false discovery rates have been considered in the realm of high-dimensional modeling, although thus far this research has focused exclusively on tail-area approaches. The majority of this work has been concentrated on lasso regression \citep{tibshirani_1996}, a popular modeling approach which naturally performs variable selection by using $L_1$ regularization. The issue of false discovery rate control is particularly important in this case, as false discoveries can be quite prevalent for lasso models \citep{Su2017}.

The false discovery rate control provided by large-scale testing approaches is marginal in the sense that a feature $X_j$ is considered a false discovery only if that feature is marginally independent of the outcome $Y$: $X_j \independent Y$. In regression, where many features are being considered simultaneously, the issue is more complicated and can involve various kinds of conditional independence. For example, we can adopt a \textit{fully conditional} perspective, which considers a feature $X_j$ to be a false discovery if it is independent of the outcome conditional upon all other features: $X_j \independent Y | X_{k \neq j}$. Several approaches to controlling the fully conditional false discovery rate have been proposed, including procedures based on the bootstrap \citep{Dezeure2017}, de-biasing \citep{Javanmard2014}, sample splitting \citep{Wasserman2009,Meinshausen2009}, and the knock-off filter \citep{barber2015, candes2018}.

An alternative for penalized models is the \textit{pathwise conditional} perspective. Pathwise approaches focus on the point in the regularization path at which feature $j$ first becomes active and condition only on the other variables present in the model (denote this set $M_j$) when assessing whether or not variable $j$ is a false discovery: $X_j \independent Y | M_j$. The methods of \citet{CovTest} and \citet{Selective_Inference} used in conjunction with the sequential stopping rule of \citet{GSell2016} allow for control over the pathwise conditional Fdr. 

Less restrictive approaches to false discovery rates for penalized regression models have also been proposed. \citet{Breheny2019} developed an analytic method which bounds the \textit{marginal} false discovery rate (mFdr) of penalized linear regression models. \citet{Miller2019} extended this approach to a more general class of penalized likelihood models, while \citet{HuangFDR} addressed a similar question using a Monte Carlo approach.

In this paper we combine the marginal perspective on false discoveries employed by \citet{Breheny2019} and \citet{Miller2019} with the idea of local false discovery rates. The resulting method provides a powerful new inferential tool for penalized regression models, allowing one to assess each individual feature's probability of having been selected into the model by chance alone.

\subsection{Large-scale testing, Fdr, and fdr}

Consider data of the usual form $(\y, \X)$, where $\y$ denotes the response for $i = \{1, \ldots, n\}$ independent observations and $\X$ is an $n$ by $p$ matrix containing the values of $j = \{1, \ldots, p\}$ explanatory features. We presume that only a subset of the available features have a non-null relationship with the outcome, and the goal of our analysis is to correctly identify those important features.

Let $\beta_j$ denote the effect of feature $j$ in relation to $\y$; this paper focuses on regression coefficients, but $\beta_j$ could also represent a difference in means or some other measure. Large-scale univariate testing considers $p$ separate null hypotheses, $H_j:\beta_j = 0$, each corresponding to a single feature, and conducts a univariate test on each of those hypotheses. These tests are usually performed by scaling the estimated effects $\{\bh_1, \bh_2, \ldots, \bh_p\}$ by their standard errors $\{\hat{s}_1, \hat{s}_2, \ldots, \hat{s}_p\}$ to obtain test statistics $\{t_1, t_2, \ldots, t_p\}$ which are used to calculate p-values $\{p_1, p_2, \ldots, p_p\}$. Alternatively, these test statistics can be converted to $z$-values defined by $z_j = \Phi^{-1}(F_t(t_j))$, where $\Phi$ is the standard normal CDF, so that the $z$-values for features with true null hypotheses will follow a N(0,1) distribution.

The false discovery rate is a tool for meaningfully aggregating the results of these tests while quantifying the expected proportion of false positives. In this paper, our primary focus is on local false discovery rates, or estimates of $\Pr(H_j | \bh_j, \hat{s}_j) = \Pr(H_j | z_j)$, the probability that feature $j$ has a null relationship with the outcome given the observed data.

A natural framework for estimating this probability is to assume that features arise from two classes with prior probability $\pi_0$ and $\pi_1$, with $f_0(z)$ denoting the density of the $z$-values for features in the null class and $f_1(z)$ denoting the density for features in the non-null class, with the mixture density $f(z) = \pi_0f_0(z) + \pi_1f_1(z)$ giving the marginal distribution of $z$. In addition, let $\cZ$ denote any subset of the real line, $F_0(\cZ) = \int_{\cZ}^{}f_0(z)dz$ denote the probability of observing $z \in \cZ$ for a null feature and $F_1(\cZ) = \int_{\cZ}^{}f_1(z)dz$ the probability for a non-null feature

Given the framework described in the preceding paragraph, applying Bayes' rule yields
\begin{align} \label{eq:FdrBayes}
\Pr(\text{Null}|z \in \cZ) = \frac{\pi_0F_0(\cZ)}{F(\cZ)} = \Fdr(\cZ),
\end{align}
where $F(\cZ) = \pi_0F_0(\cZ) + \pi_1F_1(\cZ)$. In a typical application, $\cZ$ would denote a tail area such as $z > 3$ or $\abs{z} > 3$, and would allow the analyst to control the Fdr through the choice of $\cZ$.

Instead of focusing on the tail area condition $z \in \cZ$, we may alternatively consider the limit as $\cZ$ approaches the single point $z_j$, in which case the distribution functions become density functions and we have
\begin{align} \label{eq:fdrzch2}
\Pr(\text{Null}|z = z_j) = \frac{\pi_0f_0(z_j)}{f(z_j)} = \fdr(z_j).
\end{align}
This quantity is typically referred to as the ``local'' false discovery rate, and describes feature $j$ specifically, as opposed to the collection of features whose $z$-statistics fall in $\cZ$.

There are several important connections between Fdr and fdr \citep{Efron_locfdr}. Of particular interest is the relationship
\begin{equation} \label{eq:Fdr-fdr}
\Fdr(\cZ) = \Ex\big(\fdr(z)| z \in \cZ\big).
\end{equation}
In words, the Fdr of the set of features whose normalized test statistics fall in the tail region $\cZ$ is equal to the average fdr of the features in $\cZ$. This ensures that selecting individual features using a threshold $fdr(z) < \alpha$ also limits Fdr below $\alpha$ for the entire set of features defined by the threshold. Additional detail on the links between Fdr and fdr can be found in \citet{Efron_locfdr} and \citet{Strimmer2008}.

Broadly speaking, there are two main approaches to estimating local false discovery rates given the observed collection $\{z_j\}_{j=1}^p$. Recall that by construction, $f_0$ is the density function of the standard normal distribution; thus, $\pi_0$ and $f$ are the only quantities that must be estimated. The first approach, originally proposed in \citep{Efron_locfdr} but extended and modified by many authors since then, focuses on estimating the marginal density $f$. Replacing $\pi_0$ with its upper bound of 1, and estimating $f(z)$ using any nonparametric density estimation method (e.g., kernel density estimation), we have
\begin{align} \label{eq:fdrhat1}
\widehat{\fdr}(z_j) = \frac{f_0(z_j)}{\hat{f}(z_j)}.
\end{align}

Alternatively, one can explicitly model the mixture distribution of $z$ (in this approach, there is typically one null distribution and many non-null distributions), obtaining the estimates $\hat{\pi}_0, \hat{\pi}_1, \hat{\pi}_2, \ldots$. The estimated local fdr is therefore
\begin{align} \label{eq:fdrhat2}
\widehat{\fdr}(z_j) = \frac{\hat{\pi}_0f_0(z_j)}{\sum_{k=0}^K\hat{\pi}_k\hat{f}_k(z_j)},
\end{align}
where $K$ is the number of non-null mixture components and $\hat{f}$ must be estimated for the non-null components; estimation of $\pi_k$ and $f_k$ is typically accomplished with maximum likelihood via EM algorithm. This approach was originally proposed by \citet{Muralidharan2010}, but as with the marginal density approach, has been explored by many other authors since then. In particular, we focus on a mixture model proposed by \citet{Stephens2017}, which requires that all non-null components have a mode of 0 (Stephens refers to this as the ``unimodal assumption''). One attractive aspect of this model, which is implemented in the R package \texttt{ashr}, is that the resulting $fdr$ is a monotone function of as $z$ moves away from 0 in either direction; this is typically not true for marginal density estimates of the form \eqref{eq:fdrhat1}.

In the sections that follow we refer to fdr estimates based on \eqref{eq:fdrhat2} as the ``ashr'' approach, and fdr estimates found using \eqref{eq:fdrzch2} as the ``density'' approach. Both ashr and density approaches estimate the posterior probability, given the observed data, that feature $j$ is a false discovery, but their assumptions can lead to different estimates. We further discuss the relative strengths and weaknesses of these two approaches in Section~\ref{Sec:density_est}.

\subsection{Penalized regression and mFdr}
\label{Sec:background_mfdr}

In contrast with the univariate nature of large-scale testing, regression models simultaneously relate all of the explanatory features in $\X$ with $\y$ using a probability model involving coefficients $\bb$. In what follows we assume the columns of $\X$ are standardized such that each variable has a mean of $0$ and $\sum_i \x_{ij}^2 = n$. The fit of a regression model can be summarized using the log-likelihood, which we denote $\ell(\bb|\X,\y)$. In the classical setting, $\bb$ is estimated by maximizing $l(\bb|\X,\y)$. However, this approach is unstable when $p > n$ unless an appropriate penalty is imposed on the size of $\bb$.
In the case of the lasso penalty, estimates of $\bb$ are found by minimizing the objective function:
\al{eq:obj}{
Q(\bb|X,\y) = -\frac{1}{n} \ell(\bb|X,\y) + \lambda||\bb||_1
}

The maximum likelihood estimate is found by setting the score, $\u(\bb) = \nabla \ell(\bb|\X,\y)$, equal to zero. The lasso estimate, $\bbh$, can be found similarly, although allowances must be made for the fact that the penalty function is typically not differentiable. These penalized score equations are known as the Karush-Kuhn-Tucker (KKT) conditions in the convex optimization literature, and are both necessary and sufficient for a solution $\bbh$ to minimize $Q(\bb|\X,\y)$. 

An important property of the lasso is that it naturally performs variable selection. The lasso estimates are sparse, meaning that $\bh_j = 0$ for a large number of features, with only a subset of the available features being active in the model. The regularization parameter $\lambda$ governs the degree of sparsity with smaller values of $\lambda$ leading to more variables having non-zero coefficients.

The KKT conditions can be used to to develop an upper bound for the number of features expected to be selected in a the lasso model by random chance. Heuristically, if feature $j$ is marginally independent of $\y$, then $Pr(\bh_j \neq 0)$ is approximately equal to $Pr(\tfrac{1}{n}\abs{u_j(\bb)} > \lam)$, where $\u_j$ denotes the $j^{th}$ component of the score function (gradient of the log-likelihood). Classical likelihood theory provides asymptotic normality results and allows for estimation of this tail probability, which in turn provides a bound on the mFdr. For additional details and proofs, see \citet{Miller2019}.

This approach provides an overall assessment of model selection, but it does not offer any specific information about individual features.
It is often the case that among the selected features, some appear to be clearly related to the outcome while others are of borderline significance.
For example, as suggested by \eqref{eq:Fdr-fdr}, we may select two features, one with a 1\% fdr and the other with a 39\% fdr, but the overall Fdr of the model is 20\%. Providing this level of feature-specific inference is the major motivation for estimating local false discovery rates for penalized regression.

The ability to provide feature-specific false discovery rates also allows one to overcome the tension between predictive accuracy and selection reliability. For lasso models, it is typically the case that the number of features that can be selected under Fdr restrictions is much smaller than the number of active features in model that achieves maximum predictive performance as determined by cross-validation.
This poses something of a dilemma, as we must choose between a model with sub-optimal predictions and one with a high proportion of false discoveries.
Local fdr, however, allows us to use the most predictive model while retaining the ability to identify features that are unlikely to be false discoveries.

\section{Estimating mfdr}
\label{Sec:method}

We begin by mentioning that the elements of $\bbh$ are not directly suitable for local false discovery rate estimation. In particular, for most choices of $\lam$, $\bh_j$ is exactly zero for many features, making it impossible to construct statistics with a N(0,1) distribution under the null. Instead, we use the KKT conditions, which mathematically characterize feature selection at a given value of $\lambda$, to construct normally distributed statistics appropriate for the given model. Section~\ref{Sec:linear} addresses linear regression, while Section~\ref{Sec:glm_cox} addresses GLM and Cox regression models.

\subsection{Linear regression} 
\label{Sec:linear}

Consider the linear regression setting:
\begin{align*}
\y = \X \bb + \pmb{\epsilon}, \qquad \epsilon_i \sim N(0, \sigma^2).
\end{align*}
As mentioned in Section~\ref{Sec:background_mfdr}, the lasso solution, $\bbh$, is mathematically characterized by the KKT conditions, which are given by \citep{lasso_kkt}:
\begin{alignat*}{2}
\frac{1}{n}\x_j^T(\y - \X\bbh) &= \lambda \textrm{ sign}(\bh_j) \qquad & & \text{for all } \bh_j \ne 0 \\
\frac{1}{n}\x_j^T(\y - \X\bbh) &\leq \lambda & & \text{for all } \bh_j = 0.
\end{alignat*}

We define the partial residual as $\rj = \y - \X_{-j}\bbh_{-j}$ where the subscript $-j$ indicates the removal of the $j^{th}$ feature. Using this definition it follows directly from the KKT conditions that:
\begin{alignat*}{2}
\frac{1}{n}|\x_j^T \rj| &> \lambda \qquad && \text{for all } \bh_j \ne 0 \\
\frac{1}{n}|\x_j^T \rj| &\leq \lambda && \text{for all } \bh_j = 0.
\end{alignat*}
The quantity $\frac{1}{n}\x_j^T \rj$ governs the selection of the $j^{th}$ feature: if its absolute value is large enough, relative to $\lambda$, feature $j$ is selected.
In this manner, $\frac{1}{n}\x_j^T \rj$ can be considered analogous to a test statistic in the hypothesis testing framework.

In the special case of orthonormal design where $\frac{1}{n}\X^T \X = \I$, it is straightforward to show that $\frac{1}{n}\x_j^T \rj \sim N(\beta_j, \sigma^2/n)$ \citep{Breheny2019}. Under the null hypothesis that $\beta_j = 0$, this result can be used to construct the normalized test statistic
\begin{align} \label{eq:fdr_linearch2}
z_j = \frac{\frac{1}{n}\x_j^T \rj}{\hat{\sigma}/\sqrt{n}},
\end{align}
where $\hat{\sigma}$ is an estimate of $\sigma$; for the sake of simplicity, we use the residual sum of squares divided by the model degrees of freedom \citep{zou2007}, but many other possibilities exist \citep{reid2016}. These statistics are then used to estimate local false discovery rates using either \eqref{eq:fdrhat1} or \eqref{eq:fdrhat2}.

In practice, the design matrix will not be orthonormal and the result $\frac{1}{n}\x_j^T \rj \sim N(\beta_j, \sigma^2/n)$ will not hold exactly. Nevertheless, it still holds approximately under reasonable conditions. To understand these conditions we explore the relationship between $\frac{1}{n}\X^T \X$ and $z_j$:
\begin{equation}
\begin{aligned}
\label{eq:remainder}
\frac{1}{n}\x_j^T \rj &= \frac{1}{n}\x_j^T (\X\bb + \bep - \X_{-j}\bbh_{-j}) \\
&= \frac{1}{n}\x_j^T\bep + \beta_j + \frac{1}{n}\x_j^T \X_{-j} (\bb_{-j} - \bbh_{-j}).
\end{aligned}
\end{equation}
The component $\frac{1}{n}\x_j^T\bep + \beta_j$ is unaffected by the structure of $\frac{1}{n}\X^T \X$; thus the estimator in \eqref{eq:fdr_linearch2} will be accurate in situations where the final term, $\frac{1}{n}\x_j^T \X_{-j} (\bb_{-j} - \bbh_{-j})$, is negligible (in orthonormal designs, this term is exactly zero). If feature $j$ is independent of all other features, then $\frac{1}{n}\x_j^T \X_{-j}$ will converge to zero as $n$ increases, making the term asymptotically negligible provided $\sqrt{n}(\bb_{-j} - \bbh_{-j})$ is bounded in probability.

If pairwise correlations exist between features, $\frac{1}{n}\x_j^T \X_{-j}$ will not converge to zero, and the null distribution of $z_j$ will not follow a standard normal distribution; in particular, as discussed in \citet{Breheny2019}, its distribution will have thinner tails than a standard normal distribution. This will causes the local mfdr estimates to be somewhat conservative in the presence of strong correlation; this phenomenon is explored in depth in Section~\ref{Sec:loc_sims}.

When $\frac{1}{n}\x_j^T \rj \sim N(\beta_j, \sigma^2/n)$ holds exactly, the mfdr estimator of \eqref{eq:fdr_linearch2} shares an important relationship with the mFdr estimator proposed by \citet{Breheny2019}, captured in the following theorem, whose proof appears in the appendix:

\begin{theorem}
\label{Thm:Efdr}
Let $\cM_{\lam}$ denote the set of nonzero coefficients selected by a lasso model, and let $C_j = n^{-1}\x_j\rj$ denote the random variable governing the selection of a given feature.  If $C_j$ has density $g = \pi_0 g_0 + (1 - \pi_0) g_1$, where $g_0$ is the $\Norm(0, \sigma^2/n)$ density, then
\begin{align*}
\mFDR(\cM_\lambda) = \EX\left\{\mfdr(\tfrac{C_j}{\sigma/\sqrt{n}}) | j \in \cM_\lambda\right\}.
\end{align*}
\end{theorem}

Noting that $z_j$ from \eqref{eq:fdr_linearch2} is a standardized version of $c_j$, dividing by $\sigma/\sqrt{n}$ in order to have unit variance, Theorem~\ref{Thm:Efdr} states that, on average, the marginal false discovery rate of a model is the average local false discovery rate of its selections. Alternatively, this result implies that the expected number of false discoveries in a model can be decomposed into the sum of each selected feature's mfdr.

The above theorem assumes known values for $\pi_0$, $\sigma$, and $f$; when these quantities are estimated from the data, the equality no longer holds.  In our experience, the average mfdr is typically close to the mFdr, although this depends on how the above quantities are estimated.

\subsection{GLM and Cox models}
\label{Sec:glm_cox}

We now consider the more general case where $\y$ need not be normally distributed. Specifically we focus our attention on binary outcomes (logistic regression) and survival outcomes (Cox regression), although the approach is general and can also be applied to other likelihood based models.

Similar to the linear regression setting, we can develop a local false discovery rate estimator by studying minimization of the objective function, $Q(\bb|X,\y)$, as defined in \eqref{eq:obj}. When $\y$ is not normally distributed, $\ell(\bb|X,\y)$ is no longer a quadratic function. However, we can construct a quadratic approximation by taking a Taylor series expansion of $\ell(\bb|X,\y)$ about a point $\bbt$. In the context of this approach it is useful to work in terms of the linear predictor $\be = \X\bb$ (and $\tilde{\be} = \X\bbt$), noting that we can equivalently express the likelihood in terms of $\be$ such that:
\begin{align*}
\ell(\bb|X,\y) & \approx l(\bbt) + (\bb - \bbt)^T l'(\bbt) + \frac{1}{2}(\bb - \bbt)^T l''(\bbt) (\bb - \bbt) \\
& = \frac{1}{2}(\yt - \be)^T f''(\bet) (\yt - \be) + \text{const}.
\end{align*}
Here $\yt = \bet - f''(\bet)^{-1} f'(\bet)$ serves as a pseudo-response in the weighted least squares expression.
The KKT conditions here are very similar to those in the linear regression setting, differing only by the inclusion of a weight matrix $\W = f''(\bet)$ and the fact that $\y$ has been replaced by $\yt$. 

Proceeding similarly to Section~\ref{Sec:linear}, we define the partial pseudo-residual $\rj = \yt - \X_{-j}\bbh_{-j}$, which implies:
\begin{alignat*}{2}
\frac{1}{n}|\x_j^T \W \rj| &> \lambda \qquad & &\text{for all } \bh_j \ne 0 \\
\frac{1}{n}|\x_j^T \W \rj| &\leq \lambda & &\text{for all } \bh_j = 0.
\end{alignat*}
\citet{Miller2019} show that, under appropriate regularity conditions,
\begin{align}\label{eq:glm_fdr}
z_j=\frac{\tfrac{1}{n}\x_j^T\W\rj}{\hat{s}_j/\sqrt{n}} \inD N(0, 1),
\end{align}
where $\hat{s}_j = \sqrt{\x_j^T\W\x_j/n}$. As in Section~\ref{Sec:linear}, these statistics can be used to estimate local false discovery rates using either \eqref{eq:fdrhat1} or \eqref{eq:fdrhat2}.

For the most part, the regularity conditions required for \eqref{eq:glm_fdr} to hold are the same as those required for asymptotic normality in classical likelihood theory, with one additional requirement. Just as the mfdr estimator in Section~\ref{Sec:linear} holds under feature independence, the estimator in Equation~\ref{eq:glm_fdr} relies on an assumption of vanishing correlation; consequently it will be accurate in the case of independent features but tends to result in conservative false discovery rate estimates when features are correlated. The assumption of vanishing correlation is unlikely to be literally true in practice; its purpose is to establish a hypothetical worst-case scenario in terms of false discovery selection so that a fdr can be estimated. We investigate how robust the estimator is to this assumption in Section~\ref{Sec:loc_sims}.

\section{Simulation studies}
\label{Sec:loc_sims}

In this section we conduct a series of simulations studying the behavior of the mfdr estimators resulting from \eqref{eq:fdr_linearch2} and \eqref{eq:glm_fdr}. We investigate both the internal validity of the method in terms of whether the estimate accurately reflects the probably that a feature is a false discovery as well as compare the results of mfdr-based inference with other approaches to inference in the high-dimensional setting.

For the mfdr approach, we present results for two different values of $\lambda$. The first, $\lambda_{\CV}$, characterizes the model with the lowest cross-validated error. The second, $\lambda_{1\SE}$, characterizes the most parsimous model within one standard error of the lowest cross-validated error. Unless otherwise indicated, we estimate mfdr using the approach of \citet{Stephens2017} as implemented in the R package \texttt{ashr}.

We generate data from three models: linear regression, logistic regression, and Cox regression. For each model we present results for two data-generating scenarios we refer to as ``Assumptions Met'', where the underlying assumption of independent features and furthermore $n > p$, which improves asymptotic approximations, and ``Assumptions Violated'', where features are correlated in a manner consistent with real data and $p > n$.

\textbf{Assumptions Met:} In this scenario, $n > p$ and all features are independent of each other. Both of these factors should lead to a small remainder term in \eqref{eq:remainder}.

\begin{itemize}
\item $n = 1000, p=600$
\item Covariate values $x_{ij}$ independently generated from the standard normal distribution.
\item Response variables are generated as follows:
\begin{itemize}
\item Linear regression, $\y = \X\bb + \bep$ where $\epsilon_i \sim N(0,\sigma^2)$, $\bb_{1:60} = 4$, and $\bb_{61:600} = 0$, and $\sigma = \sqrt{n}$
\item Logistic regression, $y_i \sim \textrm{Bin}\bigg(1, \pi_i = \frac{\exp(\x_i^T \bb)} {1 + \exp(\x_i^T \bb)}\bigg)$, $\bb_{1:60} = .15$, and $\bb_{61:600} = 0$
\item Cox regression, $y_i \sim \textrm{Exp}\big(\exp(\x_i^T \bb)\big)$, $\bb_{1:60} = .15$, and $\bb_{61:600} = 0$, and 10\% random censoring
\end{itemize}
\end{itemize}


\textbf{Assumptions Violated:} In this scenario, we impose an association structure motivated by the causal diagram below.
\begin{center}
\begin{tikzpicture}[node distance=1cm]

\node(b)[text centered] {$B$};
\node(u)[below of = b, text centered] {$ $};
\node(a)[left of = u, text centered, xshift = -1.5cm] {$A$};
\node(c)[right of = u, text centered, xshift = 1.5cm] {$C$};
\node(y)[below of = u, text centered] {$Y$};
\draw [arrow] (a) -- (b);
\draw [arrow] (a) -- (y);

\end{tikzpicture} \\
\end{center}
Here, variable $A$ has a direct causal relationship with the outcome variable $Y$, variable $B$ is correlated with $Y$ through its relationship with $A$, but is not causally related, and variable $C$ is unrelated to all of the other variables and the outcome. In terms of the false discovery perspectives introduced in Section~\ref{Sec:mfdr_back}, all of the perspectives agree that $A$ would never be a false discovery and that $C$ would always be a false discovery. However, selecting $B$ is considered a false discovery by the fully conditional perspective, but not the marginal perspective. From the pathwise conditional perspective, whether $B$ is a false discovery depends on whether $A$ has entered the model or not.

\begin{itemize}
\item $n = 200, p=600$
\item Covariates generated with the following dependence structure:
\begin{itemize}
\item 6 causative features ($A$), which are independent of each other
\item 54 correlated features ($B$), grouped such that 9 are related to each causative feature with $\rho = 0.5$
\item 540 noise features ($C$), which are correlated with each other by an autoregressive correlation structure where $\textrm{Cor}(\x_j, \x_k) = 0.8^{|j - k|}$
\end{itemize}
\item Response variables ($Y$) are generated from the same models described in the Assumptions Met scenario; however, $\bb$ differs to reflect the change in sample size:
\begin{itemize}
\item Linear regression, $\bb_{1:6} = (6,-6,5,-5,4,-4)$, and $\bb_{7:600} = 0$, and $\sigma = \sqrt{n}$
\item Logistic regression, $\bb_{1:6} = (1.1,-1.1,1,-1,.9,-.9)$, and $\bb_{7:600} = 0$
\item Cox regression, $\bb_{1:6} = (.6, -.6, .5, -.5, .4, -.4)$, and $\bb_{7:600} = 0$, and 10\% random censoring
\end{itemize}
\end{itemize}

To summarize each of these scenarios in terms of the diagram above, the Assumptions Met scenario consists of 60 features akin to variable $A$, 0 features akin to variable $B$, and 540 features akin to variable $C$, while the Assumptions Violated scenario consists of 6 features akin to variable $A$, 54 features akin to variable $B$, and 540 features akin to variable $C$. The Assumptions Violated scenario also imposes an autoregressive correlation structure on the noise features, undermining the assumption of vanishing correlation.

For comparison, we also include results for the traditional univariate approach to fdr throughout our simulations, defining the univariate procedure to consist of fitting a univariate regression model to each of the $j \in \{1, \ldots, p\}$ features, extracting the test statistic, $t_j$, corresponding to the test on the hypothesis that $\beta_j = 0$, then normalizing these test statistics such that $z_j = \Phi^{-1}(Pr(T < t_j))$. The \texttt{ashr} package was then used to calculate local false discovery rates.

\subsection{Calibration}

How well do our proposed estimates reflect the true probability that a feature is purely noise (i.e., unrelated to the outcome either directly or indirectly)? We address this question through calibration plots comparing our mfdr estimates to the observed proportion of noise features across the full spectrum of false discovery rates. For example, an estimate of $0.2$ is well-calibrated if 20\% of features with $\widehat{\text{mfdr}} = .2$ are observed to be false discoveries. 

\begin{figure}[hbt]
\centering
\includegraphics[width=.85\textwidth]{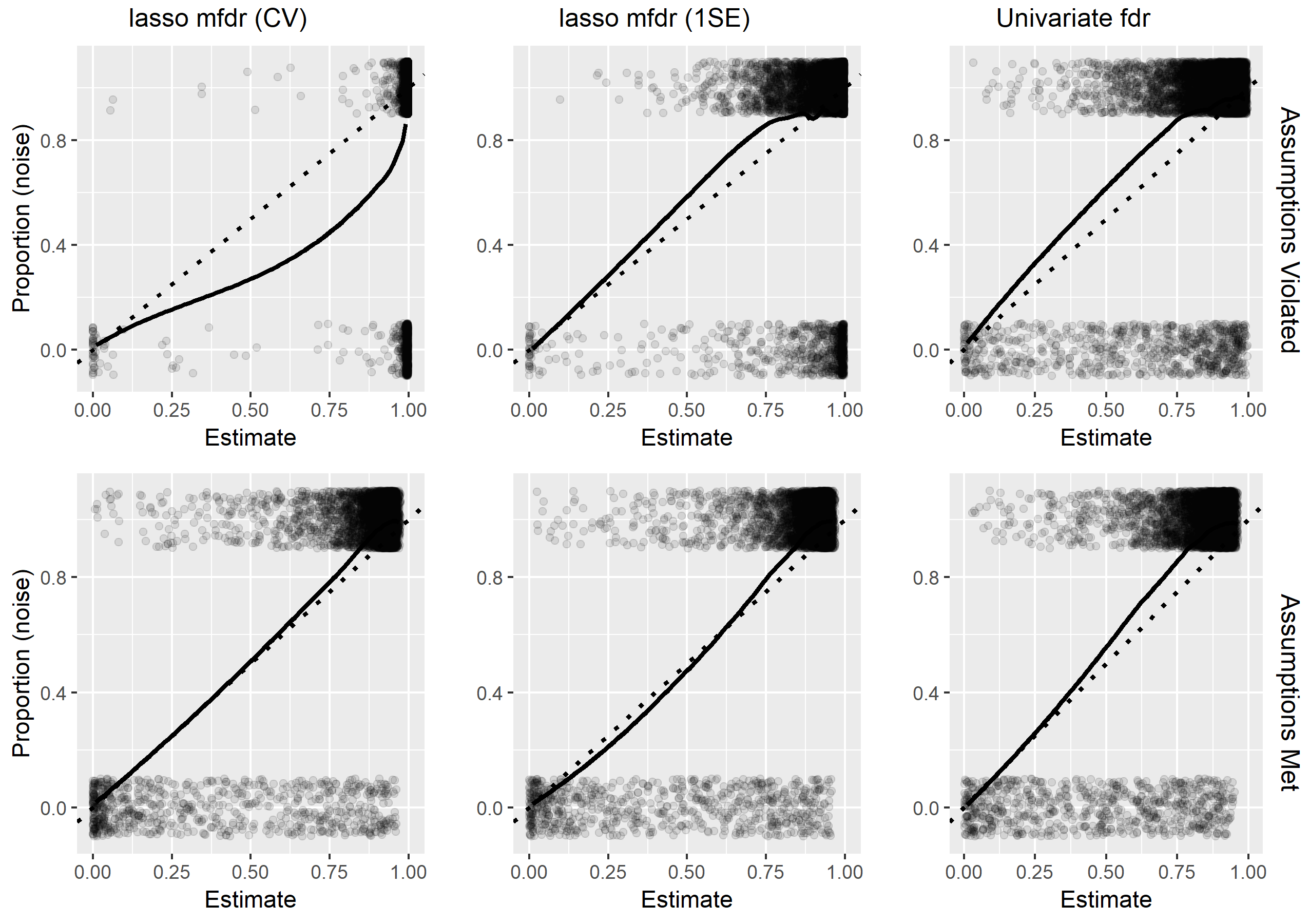}
\caption{\label{Fig:Calibr} The expected proportion of false discoveries at a given estimated local false discovery rate after smoothing for the linear regression setting. Univariate estimates along with two lasso estimates (at different values of $\lam$) are shown.
}
\end{figure}

Figure~\ref{Fig:Calibr} displays results for the linear regression setting for both scenarios.
Strip plots of individual mfdr estimates are provided along with smoothed estimates of the calibration relationships and a 45 degree line is provided for reference.
Similar results are available in Supplemental Material for logistic and Cox models; the same patterns hold for all three models.

When all assumptions are met, the mfdr estimates are very accurate at both $\lambda_{\CV}$ and $\lambda_{1\SE}$, showing essentially perfect correspondence with the empirical proportion of false discoveries. The traditional univariate method also appears to be very well calibrated, only slightly underestimating the proportion of noise features at the upper end of the mfdr spectrum.

In the Assumptions Violated scenario, the lasso mfdr estimates tend to be accurate in the regions near 0 and 1, which are typically of greatest practical interest. In between, the estimates based on $\lambda_{\CV}$ are conservative.  For example, among features with an estimated mfdr of 35\%, only about 20\% were actually noise features.  The lasso mfdr estimates based on $\lambda_{1\SE}$ and the univariate estimates, on the other hand, slightly underestimated the probability that a given feature was a false discovery.

It is worth noting that the presence of correlated ``B'' variables in the Violated scenario complicates the assessment of calibration, as features cannot be unambiguously separated into noise and signal. Since our method is based upon the marginal perspective, we do not treat the ``B'' variables as false (noise) discoveries here. However, as one would expect, they tend to have much higher mfdr estimates than the causative (``A'') variables; this should be kept in mind when interpreting these calibration plots.

The three methods depicted in Figure~\ref{Fig:Calibr} show markedly different potential with respect to classifying features as noise versus signal. At $\lambda_{\CV}$, the mfdr estimates of noise features are tightly clustered near 1 and the mfdr estimates of causal features are tightly clustered near 0, as seen in the strip plots at the top and bottom of the figure, respectively. In contrast, the traditional univariate method yields far more intermediate estimates for variables of both types. In other words, the lasso estimates allow one to much more confidently identify signals compared to a univariate analysis.  As expected, the results for $\lambda_{1\SE}$ fall somewhere in between the results at $\lambda_{\CV}$ and the univariate approach.

\begin{table}[!htb]
\centering
\caption{\label{Table:hard} Local false discovery rate accuracy results for the Assumptions Violated scenario. Features are binned based upon their estimated fdr and the observed proportion of noise variables in each bin is reported in the body of the table for each method.}
\begin{tabular}{ l c c c c c }
\hline
Linear & (0, 0.2] & (0.2, 0.4] & (0.4, 0.6] & (0.6, 0.8] & (0.8, 1] \\
\hline
Univariate $\widehat{\text{fdr}}$ & 0.11 & 0.40 & 0.63 & 0.85 & 0.95 \\ 
$\widehat{\mfdr}$ at $\lambda_{1\SE}$ & 0.03 & 0.36 & 0.60 & 0.84 & 0.92 \\ 
$\widehat{\mfdr}$ at $\lambda_{\CV}$ & 0.03 & 0.25 & 0.32 & 0.49 & 0.91 \\ 
\hline
Logistic & (0, 0.2] & (0.2, 0.4] & (0.4, 0.6] & (0.6, 0.8] & (0.8, 1] \\
\hline
Univariate $\widehat{\text{fdr}}$ & 0.15 & 0.45 & 0.68 & 0.87 & 0.95 \\ 
$\widehat{\mfdr}$ at $\lambda_{1\SE}$ & 0.00 & 0.05 & 0.15 & 0.29 & 0.91 \\ 
$\widehat{\mfdr}$ at $\lambda_{\CV}$ & 0.00 & 0.00 & 0.07 & 0.31 & 0.91 \\ 
\hline 
Cox & (0, 0.2] & (0.2, 0.4] & (0.4, 0.6] & (0.6, 0.8] & (0.8, 1] \\
\hline
Univariate $\widehat{\text{fdr}}$ & 0.08 & 0.39 & 0.65 & 0.86 & 0.95 \\ 
$\widehat{\mfdr}$ at $\lambda_{1\SE}$ & 0.00 & 0.00 & 0.07 & 0.29 & 0.91 \\ 
$\widehat{\mfdr}$ at $\lambda_{\CV}$ & 0.00 & 0.00 & 0.00 & 0.26 & 0.91 \\
\hline
\end{tabular}
\end{table}

Table~\ref{Table:hard} displays an alternative representation of the calibration results for the Assumptions Violated scenario. Here, features are sorted into five equally spaced bins based upon their estimated local false discovery rate under a given procedure. Within in each bin we calculate the proportion of noise (i.e., ``C'') variables; for a well-calibrated estimator, the proportion of noise features within a bin should remain within the range of the bin.  For example, in the linear regression case, 25\% of the features with $\widehat{\mfdr}$ estimates between 0.2 and 0.4 at $(\lam_\CV)$ were truly noise.  Overall, the mfdr estimates at $\lam_\CV$ were always either well-calibrated or conservative, with the estimates for logistic and Cox regression more conservative than those for linear regression.  In comparison, the univariate approach occasionally underestimated the false discovery probability.  Again, results for $\lam_{1\SE}$ were intermediate between the other two approaches.

\subsection{Power compared to univariate fdr}

In this section we compare the number of mfdr feature selections of each type, $A$, $B$, and $C$, for the two previously described scenarios to the selections based on univariate local false discovery rates.

\begin{figure} [htb]
\centering
\includegraphics[width=.8\textwidth]{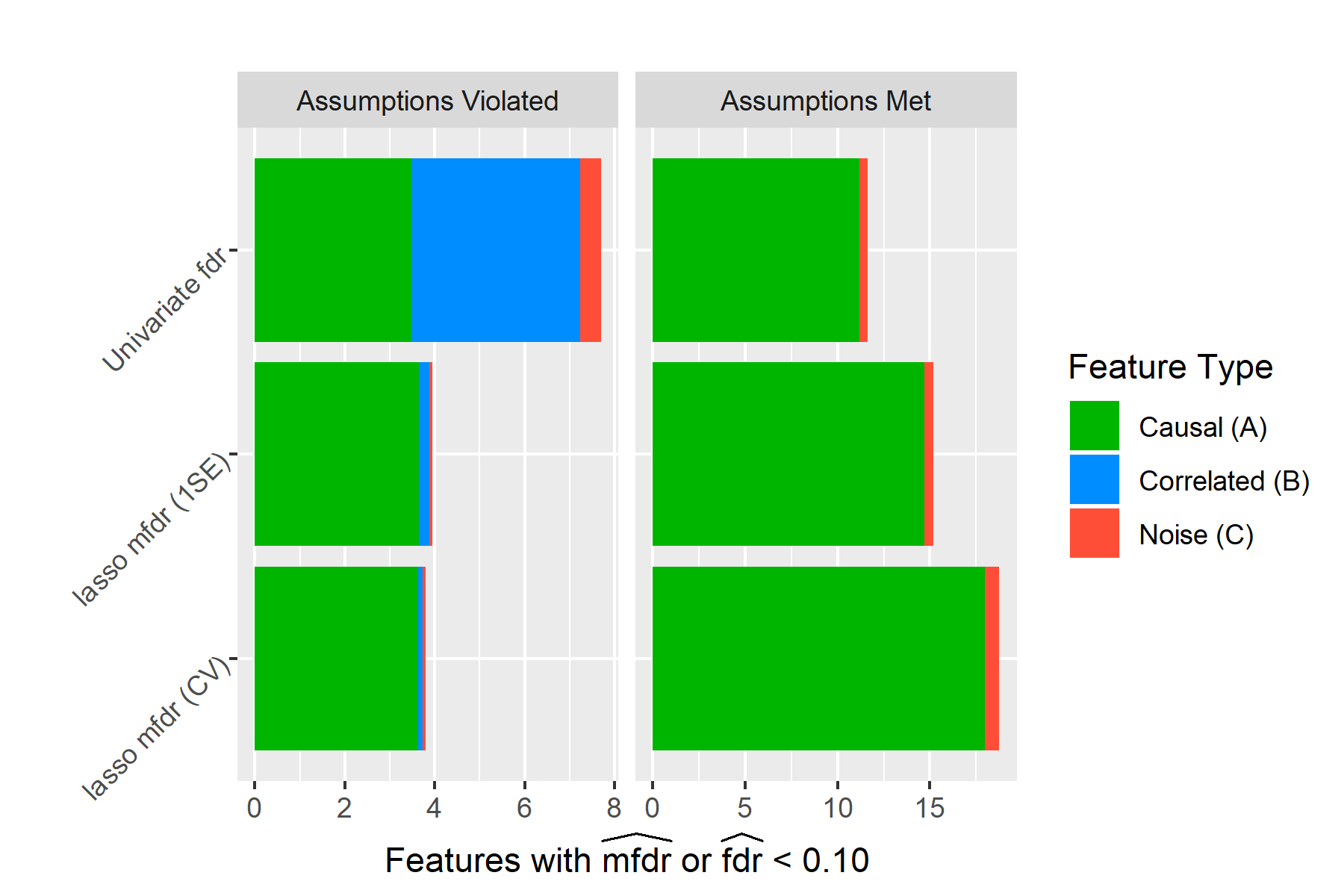}
\caption{\label{Fig:power} The average number of features of each type with estimated local false discovery rates of less than 0.10 for each method in the two scenarios.}
\end{figure}

Using a local false discovery rate threshold of $0.1$, Figure~\ref{Fig:power} shows that the regression-based mfdr approach leads to increased selection of causally important variables. At $\lambda_{\CV}$, the lasso mfdr approach selects 61\% more causal (``A'') variables than the univariate approach in the Assumptions Met scenario. In the Assumptions Violated scenario, lasso mfdr ($\lam_\CV$) remains slightly more powerful than the univariate approach, selecting on average 5\% more $A$ variables (3.63 vs. 3.45) than the univariate approach.

In addition to improving the power to detect causal variables, the mfdr approach drastically reduces the amount of correlated, non-causal features with low local false discovery rate estimates. This is most notable when comparing mfdr at $\lambda_{\CV}$ with univariate fdr, where the number of $B$ variables with fdr $< 0.1$ is \textit{36 times higher} for the univariate approach.
Furthermore, in the presence of correlated (``B'') features, a univariate approach also selects 6.6 times as many features that are purely noise (``C'') when compared to mfdr at $\lambda_{\CV}$.
Thus, the penalized regression mfdr approach proposed here results in selecting increased numbers of causally important variables while also reducing the number of correlated and noise features selected.

The results shown in Figure~\ref{Fig:power} use a somewhat arbitrary threshold of $0.1$.
To illustrate performance over the entire spectrum of classification thresholds, we also performed an ROC analysis. Specifically, we considered the number of false positives, defined as noise features classified as significant at a given threshold, and false negatives, defined as important features classified as noise at a given threshold, and assessed discriminatory power using the area under the ROC curve (AUC). For the Assumptions Violated scenario, we omit $B$ variables from these calculations.
At $\lambda_{\CV}$, the mfdr approach results in average AUC values of 0.936 and 0.990, respectively, for the Met and Violated Scenarios. This is an improvement over the average AUC values of 0.908 and 0.966 for the univariate procedure, further demonstrating the advantages of regression-based mfdr over univariate approaches.


\subsection{Comparisons with other inferential approaches for penalized regression}

Theorem~\ref{Thm:Efdr} indicates that mfdr can be used to control mFdr, motivating a comparison of the mfdr method and existing Fdr control approaches for lasso regression models. In this simulation, we use mfdr to select features with $\widehat{\mfdr} < 0.1$. This is a conservative approach to mFdr control -- recalling the relationship between mfdr and mFdr given in Theorem~\ref{Thm:Efdr}, if $\widehat{\mfdr} < 0.1$ for every feature, then the average mfdr will be $\ll 0.1$ -- but serves to illustrate the most salient differences between local mfdr and other inferential approaches.

We compare our results with the selective inference approach of \citet{Selective_Inference} using the ForwardStop rule \citep{GSell2016}, which controls the pathwise-wise Fdr at 10\%, the repeated sample splitting method as implemented by the {\tt hdi} package \citep{Dezeure2015}, which controls the fully conditional Fdr at 10\%, and the (Model-X) knock-off filter method implented in the {\tt knockoff} package \citep{candes2018}, which also controls the fully conditional Fdr at 10\%.

\begin{table}[!htb]
\centering
\caption{\label{Tab:SelectiveInference} Simulation results comparing the average number of selections of causal, correlated, and noise variables, as well as the proportion of noise variable selections, for various model-based false discovery rate control procedures. The ``exact'', ``spacing'', ``mod-spacing'', and ``covtest'' methods are related tests performed by the {\tt selectiveInference} package. Noise rate here refers to the fraction of selected variables that come from the ``Noise'' group of features (i.e., the mFdr).}
\begin{tabular}{l r r r r @{\hskip 0.5in} r r r}
& \multicolumn{4}{c}{Assumptions Violated} & \multicolumn{3}{c}{Assumptions Met}\\
\hline
& Causal & Correlated & Noise    & Noise rate & Causal  & Noise    & Noise rate\\
& (of 6) & (of 54)    & (of 540) & (mFdr)     & (of 60) & (of 540) & (mFdr)\\
\hline
mfdr (CV) & 3.84 & 0.60 & 0.14 & 3.0 \% & 16.25 & 0.6 & 4.0 \% \\
multi-split & 2.03 & 0.05 & 0 & 0\% & 10.53 & 0 & 0 \% \\
knock-off & 0.42 & 0.16 & 0.02 & 5.0 \% & 3.50 & 0 & 0\% \\
exact & 0.84 & 0.060 & 0 & 0\% & 0.84 & 0 & 0\%\\
spacing & 1.45 & 0.070 & 0 & 0 \% & 0.98 & 0 & 0\% \\
mod-spacing & 1.45 & 0.10 & 0 & 0 \% & 0.98 & 0 & 0\% \\
covtest & 1.43 & 0.10 & 0 & 0 \% & 0.97 & 0 & 0\% \\
\hline
\end{tabular}
\end{table}

Table~\ref{Tab:SelectiveInference} shows the average number of causal and correlated features selected along with the false discovery rate for the aforementioned approaches. As they are intended to, the conditional Fdr approaches greatly limit the number of correlated, non-causal (type B) features present in the model compared to the proposed marginal fdr approach. However, this added control comes at a considerable cost in terms of power to discover causal features of interest.  Overall, using pathwise or fully conditional approaches resulted in discovering at least 50\% fewer causative features, and with the conditional approaches discovering up to 95\% fewer features than the marginal approach in some cases.  These results demonstrate that conditional Fdr control approaches tend to be much more conservative than marginal approaches, even in moderate dimensions.

\subsection{Differences in the ashr and density approaches}
\label{Sec:density_est}

In Sections \ref{Sec:mfdr_back} and \ref{Sec:method}, we discussed two methods of local false discovery rate estimation, the ``ashr'' approach, which estimates fdr using the posterior distribution of feature effects under a mixture model \citep{Stephens2017}, and the ``density'' approach, which estimates fdr as the ratio of the theoretical null to an empirically estimated mixture density. It is beyond the scope of this paper to exhaustively review the differences of these approaches, but we did explore their performance in the simulations described in section~\ref{Sec:loc_sims}. Overall, the approaches provide very similar results in most cases (see Supplemental Material), although each one has strengths and weaknesses.

The primary weakness of the mixture modeling approach is that when a very high proportion of features are null (i.e., when $\pi_0 \approx 1$), there is very little information with which to estimate the non-null mixture components.  In particular, it is not uncommon for the mixture modeling approach used by \texttt{ashr} to estimate $\pi_0=1$ even when non-null features are present, especially when the signal is weak or the sample size is small.  Certainly, this does not invalidate the procedure -- one could argue that it is wise to be cautious in the presence of weak signals -- although it does contribute, in some scenarios, to the method being conservative.

\begin{figure} [htb]
\centering
\includegraphics[width=.95\textwidth]{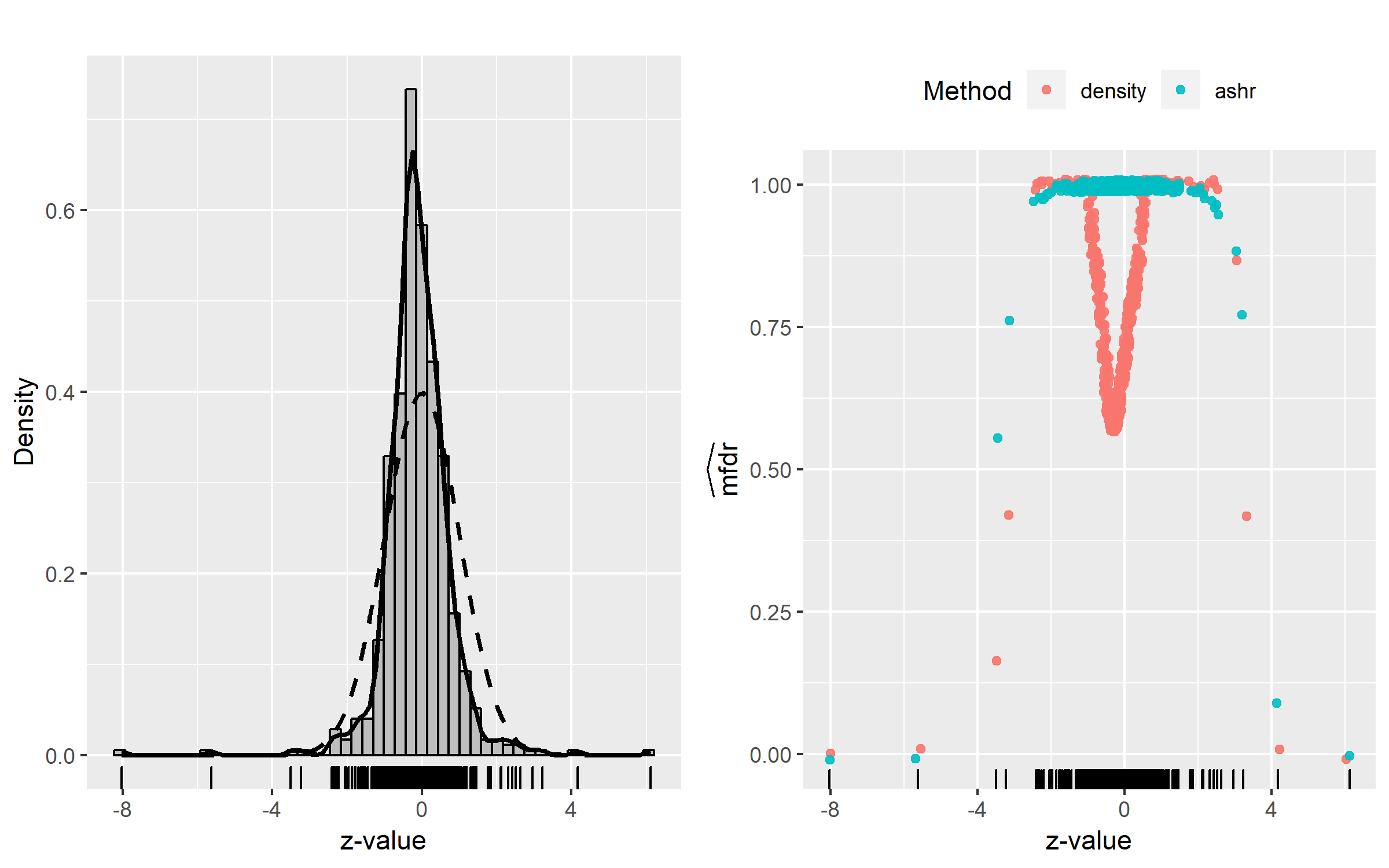}
\caption{\label{Fig:density_comparison}The relationship between feature $z$-values and their estimated local false discovery rates in the assumptions violated simulation scenario when using the ashr and density approaches. The right panel shows the empirical distribution of $z$-values (histogram and rug plot), the estimated mixture density using kernel density estimation (solid line), and the theoretical N(0,1) null distribution (dashed line). The left panel shows the relationship between $z$-values and mfdr estimates for each method.}
\end{figure}

The primary weakness of the density modeling approach is that the resulting mfdr is not a monotone function of the normalized test statistic $z$ and as a result, is somewhat prone to artifacts as shown in Figure~\ref{Fig:density_comparison}.  The figure depicts results from a single dataset simulated under the assumptions violated scenario. Due to the numerous pairwise correlations between noise features the null distribution of $z$ is more tightly concentrated around zero than the theoretical $N(0,1)$ suggests.  For the density modeling approach, this has the undesirable consequence that features with $z$-values near zero have fdr estimates smaller than those of features with $z$-values further away from zero. The mixture modeling approach of \texttt{ashr}, which uses unimodal mixture components, protects against this and leads to a monotonic relationship between a feature's local false discovery rate estimate and how far its $z$-value is from zero.

For this reason, we consider it safer in general to use the \texttt{ashr} approach and make it the default method in the \texttt{ncvreg} package (provided the \texttt{ashr} package is installed).  However, as the figure suggests, the two approaches tend to produce very similar results in the tails of the $z$ distribution.  If one is careful to avoid artifacts away from the tails, both methods provide valid results; in the next section, we present analyses of real data using both approaches.

\section{Case studies}
\label{Sec:loc_case}

\subsection{BRCA1 Gene Expression}

Our first case study examines the gene expression of breast cancer patients from The Cancer Genome Atlas (TCGA) project. The data set is publicly available at \url{http://cancergenome.nih.gov}, and consists of 17,814 gene expression measures for 536 subjects. One of these genes, BRCA1, is a tumor suppressor that plays a critical role in the development of breast cancer. When BRCA1 is under-expressed the risk of breast cancer is significantly increased, which makes genes that are related to BRCA1 expression interesting candidates for future research.

One would expect a large number of genes to have indirect relationships with BRCA1, and relatively few genes to directly affect BRCA1 expression, as in the following diagram:
\begin{center}
\begin{tikzpicture}[node distance=1cm]

\node(b)[text centered] {Correlated Gene};
\node(u)[below of = b, text centered] {$ $};
\node(a)[left of = u, text centered, xshift = -1.5cm] {Promoter/Repressor Gene};
\node(y)[below of = u, text centered] {BRCA1};
\draw [arrow] (a) -- (b);
\draw [arrow] (a) -- (y);
\draw [dashed] (b) to [bend left](y);

\end{tikzpicture} \\
\end{center}
Unsurprisingly, at an fdr threshold of $0.10$, the univariate approach selects 8,431 genes, clearly picking up on a large number of indirect associations.

Alternatively, we may use lasso regression to jointly model the relationship between BCRA1 and the remaining 17,813 genes. Here, cross validation selects a model containing 96 features. This model, however, has a high false discovery rate -- the average mfdr of these features, estimated using the ashr approach, is $0.760$. One can lower the false discovery rate by choosing a larger value of $\lam$ and selecting fewer features; for example, at $\lam_{1\SE}$, the model selects 49 features with an average mfdr of 0.026. However, this smaller model is considerably less accurate, raising the cross validation error by 11\%. Using the feature-specific inference that mfdr provides, however, we can base our analysis on the model with the greatest predictive accuracy and still identify which of the 96 features are likely to be false discoveries. In this example, 16 of those genes have local false discovery rates under 10\%.

\begin{table}[!htb]
\centering
\caption{\label{Table:BRCA1}The top 10 selected genes from the univariate approach and their local false discovery rate estimates (ashr) at each $\lambda$ value.}
\begin{tabular}{l c |c c c }
Gene & Chromosome & Univariate $\widehat{\fdr}$ & $\widehat{\mfdr}$ at $\lambda_{1\SE}$ & $\widehat{\mfdr}$ at $\lambda_{\CV}$ \\
\hline
C17orf53 & 17 &$<$0.0001 & 0.0005 & 0.31987 \\
TUBG1 & 17 &$<$0.0001 & 0.0310 & 0.3507 \\
DTL & 1 &$<$0.0001 & $<$0.0001 & $<$0.0001 \\
VPS25 & 17 &$<$0.0001 & $<$0.0001 & 0.0211 \\
TOP2A & 17 &$<$0.0001 & 0.0002 & 0.0020 \\
PSME3 & 17 &$<$0.0001& $<$0.0001 & 0.0004 \\
TUBG2 & 17 &$<$0.0001 & 0.0508 & 0.3599 \\
TIMELESS & 12 &$<$0.0001& 0.0123 & 0.3196 \\
NBR2 & 17 &$<$0.0001 & $<$0.00001 & $<$0.00001\\
CCDC43 & 17 &$<$0.0001 & 0.0169 & 0.3326 \\
\end{tabular}
\end{table}

Table~\ref{Table:BRCA1} displays the 10 genes with the lowest univariate local false discovery rates, along with their mfdr estimates at $\lambda_{1\SE}$ and $\lambda_{\CV}$ using the ashr approach. Similar results were found at both $\lambda$ values using the density approach. Many of the genes with the lowest fdr estimates according to univariate analysis have biological roles with no apparent connection to BRCA1, but are located near BRCA1 on chromosome 17 and therefore all correlated with each other: TUBG1, TUBG2, NBR2, VPS25, TOP2A, PSME3, and CCDC43.
Almost all of these genes are estimated to have much higher false discovery rates in the simultaneous regression model than in the univariate approach.
%
%

Other selections that have low local false discovery rates in both the univariate and lasso approaches have very plausible relationships with BRCA1. For example PSME3 encodes a protein that is known to interact with p53, a protein that is widely regarded as playing a crucial role in cancer formation \citep{zhang_p53}. Another example is DTL, which interacts with p21, another protein known to have a role in cancer formation \citep{DTL_p21}. These results demonstrate the potential of the mfdr approach to identify more scientifically relevant relationships by reducing the number of features only indirectly associated with the outcome.

\subsection{Lung Cancer Survival}

\citet{Shedden2008} studied the survival of 442 early-stage lung cancer subjects. Researchers collected expression data for 22,283 genes as well as information regarding several clinical covariates: age, race, gender, smoking history, cancer grade, and whether or not the subject received adjuvant chemotherapy. The goal of our analysis is to identify genetic features that are associated with survival after adjusting for the clinical covariates. 

We first analyze the data using the traditional univariate fdr approach, which is based upon the test statistics from 22,283 separate Cox regression models. Each of these models contains a single genetic feature in addition to the clinical covariates. Note that although these models contain more than one variable, we will refer to this as the ``univariate approach'' to indicate how the high-dimensional features are being treated.

We compare results from the univariate approach with the proposed local mfdr approach. Here, the clinical covariates are included in the model as unpenalized covariates along with the 22,283 features, to which a lasso penalty is applied. We consider both cross validation and the one standard error rule as methods of selecting $\lambda$ and estimate mfdr using the ``density" approach. Cross validation selects $\lambda=0.095$, while the 1SE approach suggests $\lambda=0.155$, corresponding with 43 and 1 genetic features being selected, respectively.

\begin{table}[!htb] 
\centering
\caption{\label{Tab:Shedden}Local false discovery rate estimates of the top ten features, when performing univariate testing, for the Shedden survival data.}
\begin{tabular}{l l | l l c | l l c }
Univariate fdr & & mfdr at $\lambda_{1\SE}$ & & & mfdr at $\lambda_{\CV}$ & & \\
\hline 
Feature & $\widehat{\text{fdr}}$ & Feature & $\widehat{\mfdr}$ & $\bh_j \ne 0$ & Feature & $\widehat{\mfdr}$ & $\bh_j \ne 0$ \\
\hline
ZC2HC1A & 0.0010 & FAM117A & 0.0576 & * & FAM117A & 0.0678 &* \\
FAM117A & 0.0014 & TERF1 & 0.1938 & & NUDT6 & 0.4238 & * \\
SCGB1D1 & 0.0016 & PTGER3 & 0.1951 & & RAB2A & 0.8658 &* \\
CHEK1 & 0.0022 & CDC42 & 0.1955 & & MAP1A & 0.8658 & * \\
HILPDA & 0.0027 & BHLHB9 & 0.1996 & & RHOA &0.8658 & * \\
CSRP1 & 0.0039 & NDST1 & 0.2005 & & PLP1 & 0.8658& * \\
PDPK1 & 0.0050 & CPT1A& 0.2137 & & GUK1 & 0.8658 & * \\
BSDC1 & 0.0051 & AFFX-M27830 & 0.2161 & & PREP & 0.8658 & * \\
XPNPEP1 & 0.0051 & BSDC1 & 0.2233 & & SCN7A & 0.8658 & * \\
ARHGEF2 & 0.0051 & ETV5 & 0.2269 & & BTBD1 & 0.8658 & * \\
\end{tabular}
\end{table}

Table~\ref{Tab:Shedden} displays the ten features with the lowest local false discovery rates for the univariate and lasso mfdr approaches.  Here, we present mfdr estimates based on the density modeling approach.  Using the \texttt{ashr} approach, results are very similar at $\lambda_{1\SE}$, but at $\lambda_{\CV}$, all features were estimated to have local false discovery rates near 1 using \texttt{ashr}.  As discussed in Section~\ref{Sec:density_est}, this may result from not having enough features to estimate the non-null mixture components.

We observe one feature, FAM117A, stands out in all approaches, albeit with different estimates. We also notice the estimates for the univariate fdr tend to be smaller than the mfdr approach at $\lambda_{1\SE}$, which in turn tend to be smaller than those at $\lambda_{\CV}$. This illustrates a key aspect of local mfdr in practice: although the development of the fdr estimator is concerned only with marginal false discoveries, the regression model is certainly making conditional adjustments.  At $\lam_{\max}$, the lasso mfdr and univariate fdr are equivalent, but as $\lam$ decreases and the model grows larger, more extensive conditional adjustments are being performed.

\begin{figure} [!htb]
\centering
\includegraphics[width=.85\textwidth]{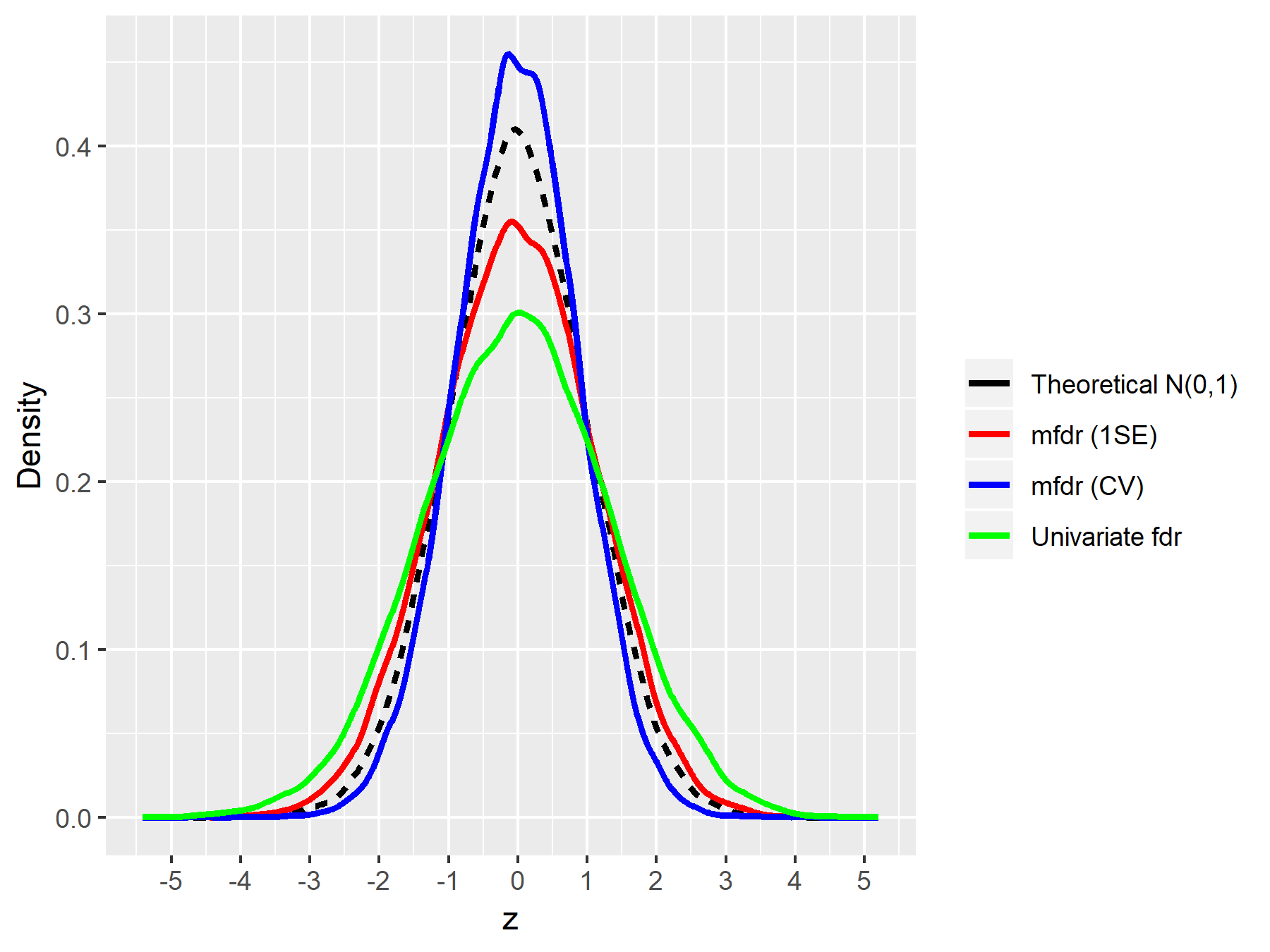}
\caption{\label{locfdr_shedden_density} The mixture density estimates, $\hat{f}(z)$, for the different methods applied to the Shedden data. We observe that the distribution of test statistics more closely resembles the null as more features are adjusted for by the model. }
\end{figure}

Figure~\ref{locfdr_shedden_density} shows the estimated marginal density, $\hat{f}$, for each method. With the univariate approach, which does not account for any correlations between features, we see that the distribution of univariate test statistics is quite different from the null distribution. In the lasso model at $\lambda_{1\SE}$ , the model adjusts for gene FAM117A and consequently, the distribution narrows relative to that of the univariate approach. When cross validation is used to select $\lambda$, the model adjusts for 43 genes and the distribution narrows even further to the point where it closely resembles the null.
As predictors enter the model and help to explain the outcome, the residuals (or pseudo-residuals) increasingly resemble white noise and exhibit no correlation with the remaining features.

\section{Discussion}

Local approaches to marginal false discovery rates for penalized regression models provides a very useful way of quantifying the reliability of individual feature selections after a model is fit. The estimator can be quickly computed, even in high dimensions, using quantities that are easily obtained from the fitted model. This makes it a convenient and informative way to carry out inference after fitting a lasso model. The method is currently implemented in the {\tt summary} function of the R package {\tt ncvreg} \citep{Breheny2011}. By default {\tt summary} reports local false discovery rates for all features selected at a given value of $\lambda$, but includes options to report all variables that meet a specified mfdr threshold or to report a specified number of features in order of mfdr significance. For more information on using {\tt ncvreg} to calculate mfdr, see the package vignette or the online documentation at \href{http://pbreheny.github.io/ncvreg}{http://pbreheny.github.io/ncvreg}.

Like any estimate, the local mfdr has limitations. Although it has clear advantages over univariate hypothesis testing in many cases, a regression approach is not practical in many situations in which high-throughput testing arises, such as two-group comparisons with $n<5$ in each group. Likewise, the local mfdr is far more powerful than other approaches to inference for the lasso such as selective inference and sample splitting, but this is because it controls a weaker notion of fdr control -- namely, it can only claim to limit the number of selections that are purely noise and does not attempt to eliminate features that are marginally associated with the outcome. Finally, although we have introduced causal ideas and diagrams to motivate ideas here, any attempt to infer causal relationships from observational data in practice should be taken with a grain of salt.

Nevertheless, the local mfdr approach that we propose here addresses a critical need for feature-specific inference in high-dimensional penalized regression models, especially in the GLM and Cox regression settings where few other options have been proposed.

\section{Appendix}
\subsection{Proof of Theorem~\ref{Thm:Efdr}}

\begin{proof}

  Let $Z=\sqrt{n}C_j/\sigma$, so that $Z$ has density $f=\pi_0 f_0 + (1-\pi_0)f_1$, with $f_0$ the standard normal density, and let $\cZ = (-\infty, -\lam\sqrt{n}/\sigma] \cup [\lam\sqrt{n}/\sigma], \infty)$, so that $Z \in \cZ$ is equivalent to $C_j \in \cM_\lam$.  Now,
  \begin{align*}
    \EX\left\{\mfdr(\tfrac{C_j}{\sigma/\sqrt{n}}) | c_j \in \cM_\lam\right\} &= \EX\left\{\mfdr(Z) | Z \in \cZ\right\} \\
    &= \int_\cZ \frac{\mfdr(z)f(z)}{F(\cZ)}\,\mathrm{d}z\\
    \intertext{(where $F(\cZ)$ denotes the probability assigned to the set $\cZ$ by the distribution function $F$)}
    &= \int_\cZ \frac{\pi_0f_0(z)}{f(z)}\frac{f(z)}{F(\cZ)} \,\mathrm{d}z\\
    &= \frac{\pi_0F_0(\cZ)}{F(\cZ)} \\
    &= \frac{2\pi_0\Phi(-\lam\sqrt{n}/\sigma)}{F(\cZ),}
  \end{align*}
  where $\Phi$ denotes the standard Gaussian CDF.  The mFdr estimator consists of replacing $F(\cZ)$ with its empirical estimate $\abs{\cM_\lam}/p$, in which case the above quantity yields the one given in Section 2.1 of \cite{Breheny2019}.
\end{proof}

\section*{Reproducibility} A repository containing code to reproduce all results in this paper is located at \url{https://github.com/remiller1450/loc-mfdr-paper}.

\bibliographystyle{ims}

\end{document}